\documentclass[fleqn,usenatbib]{mnras}

\usepackage{newtxtext,newtxmath}
\usepackage[T1]{fontenc}

\DeclareRobustCommand{\VAN}[3]{#2}
\let\VANthebibliography\thebibliography
\def\thebibliography{\DeclareRobustCommand{\VAN}[3]{##3}\VANthebibliography}

\usepackage{multirow}
\usepackage{graphicx}	% Including figure files
\usepackage{amsmath}	% Advanced maths commands
\defcitealias{2019NatAs.tmp..258I}{I19}
\defcitealias{2021MNRAS.507.1127K}{K21}
\defcitealias{2012ApJ...751....6D}{DC12}
\title[Tidal tails of $\omega$ Centauri]{Disruption of a Giant: Spectroscopic Identification of Members in the Periphery and Tidal Tails of $\omega$ Centauri\thanks{Based on observations collected at the European Southern Observatory under ESO programme 108.22MM.001.}}

\author[P. B. Kuzma et al.]{
P. B. Kuzma$^{1}$\thanks{E-mail: pete.kuzma@ed.ac.uk},
A. M. N. Ferguson$^{1}$,
A. L. Varri$^{1,2}$,
P. Bianchini$^{3}$
\\
$^{1}$Institute for Astronomy, University of Edinburgh, Royal Observatory, Blackford Hill, Edinburgh, EH9 3HJ, UK\\
$^{2}$School of Mathematics and Maxwell Institute for Mathematical Sciences, University of Edinburgh, King's Buildings, Edinburgh EH9 3FD, UK\\
$^{3}$ Université de Strasbourg, CNRS, Observatoire Astronomique de Strasbourg, F-67000 Strasbourg, France\\
}

\date{Accepted XXX. Received YYY; in original form ZZZ}

\pubyear{2023}

\begin{document}
\label{firstpage}
\pagerange{\pageref{firstpage}--\pageref{lastpage}}
\maketitle

\begin{abstract}
$\omega$ Centauri ($\omega$ Cen, NGC\,5139) is one of the most enigmatic globular clusters in the Milky Way, with the recent detection of tidal tails adding further to its complexity. We report the results of a spectroscopic study of stars in the outer regions of $\omega$ Cen, which provides an improved characterisation of the cluster periphery and confirms the existence of tidal tails. Our targets, which lie in six VLT/FLAMES fields sampling six degrees across the sky, are selected using a Bayesian inference technique.  We confirm 157 members of $\omega$ Cen based on line-of-sight velocity and [Fe/H] measurements, indicating an overall success rate of 93 per cent. We trace stars along the tidal tails to a cluster-centric radius of 3.2~deg, identifying five members in the debris and additional lower-probability candidates. The analysis of the kinematics and metallicities of the new members provides evidence of continuity in these properties from the bound component of the progenitor cluster into its tidal debris.  We find that the metallicities of stars in the peripheral regions and tidal tails of $\omega$ Cen are broadly consistent with those in the \textit{Fimbulthul} stream to which the cluster has been previously linked. Our study provides a glimpse of the promise of new and forthcoming wide-field multi-object spectrographs for advancing understanding of tidal structures around Milky Way globular clusters. 
\end{abstract}

\begin{keywords}
stars: abundances; kinematics and dynamics -- globular clusters: general -- globular clusters: individual (NGC 5139)-- Galaxy: halo.
\end{keywords}

%%%%%%%%%%%%%%%%%%%%%%%%%%%%%%%%%%%%%%%%%%%%%%%%%%

%%%%%%%%%%%%%%%%% BODY OF PAPER %%%%%%%%%%%%%%%%%%

\section{Introduction}
The origin of the most massive globular cluster (GC) in the Milky Way (MW) NGC 5139, or Omega Centauri, has been the topic of much discussion over several decades. The most popular suggestion is that Omega Centauri (hereafter $\omega$ Cen) is the nucleated core of a long-since accreted dwarf galaxy \citep[e.g.,][]{2003MNRAS.346L..11B, 2014ApJ...791..107V}, but as to which specific merger event is still unclear. Indeed, using data from {\it Gaia} Data Release 2 (DR2) \citep[][]{2018A&A...616A...1G}, $\omega$ Cen has been linked to the {\it Sequoia} \citep[][]{2019MNRAS.488.1235M} and the {\it Gaia-Enceladus-Sausage} mergers \citep[][]{2019A&A...630L...4M} based on its location in action-energy spaces, in both cases as the remnant nuclear star cluster of the progenitor. More recently, \cite{2025A&A...693A.155P} have revisited the origin of $\omega$ Cen, complementing $\it Gaia$ data with the addition of chemistry from the APOGEE survey, and argue that it stems from a potentially different merger, which they call {\it Nephele}. There are both similarities and differences between these proposed progenitor systems, which has led \cite[][]{2026arXiv260323589S} to explore the possibility that they are all part of a single merger event. However, the origin of $\omega$ Cen remains under debate.

One of the challenges in linking $\omega$ Cen to a particular merger is the complex nature of its stellar properties, which precludes using straightforward chemical tagging arguments \citep{horta2020evidence-e89}. Not only does the cluster have a broad [Fe/H] spread, ranging from --0.5 dex down to ---2.4 dex \citep[e.g.,][]{1995ApJ...447..680N,2004ApJ...605L.125B,2013MNRAS.433.1892S,2014ApJ...791..107V,2020AJ....159..254J}, but there are also quite pronounced light element variations \citep[e.g.,][]{2010ApJ...722.1373J,2017MNRAS.469..800M,2024A&A...681A..54A,2026arXiv260301041W}. Different chemical constituents are often seen to display distinct kinematic signatures; for example, \citet[][]{2018ApJ...853...86B} found that the more He-enhanced, metal-rich populations of $\omega$ Cen have a radially anisotropic velocity distribution, while the metal-poor, lower He abundance population is isotropic. Taken together, this suggests a long and complex star formation history for this cluster, though in which environment it spent most of its star-forming life is an intriguing question. \citet[][]{2002ApJ...568L..23G} argued that if $\omega$ Cen had mostly evolved in isolation (i.e., its primary form of enrichment during star formation is self-enrichment), then it is difficult to explain why it retained products from asymptotic giant branch stars while other clusters, with even higher escape speeds, did not. This provided support to the idea proposed by \citet{1993ASPC...48..608F} that the cluster is likely to be the remnant core of a disrupted dwarf. Further evidence of the extra-galactic origin of $\omega$ Cen comes from its retrograde orbit \citep[e.g.,][]{1999AJ....117.1792D,2021MNRAS.505.5978V}, making it unlikely that it formed \textit{in-situ}. If this is indeed the case, then fragments of the progenitor system should be present throughout the MW halo \citep[e.g.,][]{2018MNRAS.478.5449M, 2023MNRAS.524.2630Y}.

With the most recent data releases from the $\it Gaia$ space mission, increasingly high precision astrometry and photometry meaurements have become available for over a billion stars \citep[][]{2021A&A...649A...1G,2023A&A...674A...1G}. The astrometry measurements are particularly crucial for the identification of tenuous co-moving stellar structures in the MW halo \cite[e.g.,][]{2020MNRAS.499.2157C,2020A&A...643A..15P,2020MNRAS.495.2222S,2021ApJ...914..123I}. This is of particular relevance to $\omega$ Cen, which lies at low Galactic latitude, where identifying peripheral and tidally-stripped stars is very challenging due to the overwhelming Milky Way field populations. In the case of $\omega$ Cen, \citet[][hereafter \citetalias{2021MNRAS.507.1127K}]{2021MNRAS.507.1127K} combined photometry and astrometry from $\it Gaia$ Early Data Release 3 with a Bayesian inference model to search for faint tidal structure in its outer regions. Specifically, a mixture model was constructed that incorporated information about stellar positions, proper motions and location on the colour-magnitude diagram (CMD), and this demonstrated the existence of tidal tails, extending out to a radius of $\approx$ 450 pc ($\approx$ 5 deg). This agnostic approach enables a data-driven exploration of the regions surrounding MW GCs (Kuzma et al., in prep), and in the case of $\omega$ Cen we demonstrated how powerful this technique can be for uncovering tidal extended structures. In fact, several previous studies attempted to identify tails from $\omega$ Cen with varying degrees of success \citep[e.g.,][]{2000A&A...359..907L,2015A&A...574A..15F,2020MNRAS.495.2222S}; only with DR3 did the most compelling evidence for its tidal tails emerge.

\begin{table*}
\centering

\caption{List of observations. Columns 1) name of field (location of field along debris), 2) right ascension (J2000), 3) declination (J2000), 4) date of observation, 5) cluster-centric radius of field center in angular units, 6) cluster-centric radius of field center in units of the tidal radius ($r_{t} =  46.4$ arcmin). Where a given pointing was observed on multiple dates, both dates are listed.}
\label{tab:obs}
\begin{tabular}{cccccc}

\hline \hline
Field& R. A. & Dec. & Date& Cluster-centric&Cluster-centric\\
(location)&(J2000)&(J2000)&(dd-mm-yyyy)& radius &radius ($r_t$)\\
\hline
Outer trailing & \multirow{ 2}{*}{ 13 09 52.1}&\multirow{ 2}{*}{-45 49 42}&\multirow{ 2}{*}{28-Feb-2022}& \multirow{ 2}{*}{3.2 deg}&\multirow{ 2}{*}{4.1}\\
(OT)&&&&\\

Middle trailing & \multirow{ 2}{*}{ 13 17 37.1}&\multirow{ 2}{*}{-46 32 50}&\multirow{ 2}{*}{28-Feb-2022}& \multirow{ 2}{*}{1.8 deg}&\multirow{ 2}{*}{2.3}\\
(MT)&&&&\\

Inner trailing & \multirow{ 2}{*}{ 13 22 47.2}&\multirow{ 2}{*}{-47 04 00}&{28-Feb-2022}& \multirow{ 2}{*}{47 arcmin}&\multirow{ 2}{*}{1} \\
(IT)&&&02-Mar-2022&\\

Inner leading & \multirow{ 2}{*}{ 13 30 47.1}&\multirow{ 2}{*}{-47 52 51}&{02-Mar-2022}& \multirow{ 2}{*}{47 arcmin}&\multirow{ 2}{*}{1}\\
(IL)&&&07-Mar-2022&\\

Middle leading& \multirow{ 2}{*}{ 13 36 13.1}&\multirow{ 2}{*}{-48 21 45}&{07-Mar-2022}& \multirow{ 2}{*}{1.8 deg}&\multirow{ 2}{*}{2.3}\\
(ML)&&&30-Mar-2022&\\

Outer leading & \multirow{ 2}{*}{ 13 44 12.1}&\multirow{ 2}{*}{-48 38 12}&{02-Mar-2022}& \multirow{ 2}{*}{3.2 deg}&\multirow{ 2}{*}{4.1}\\
(OL)&&&30-Mar-2022&\\
\hline

\end{tabular}
\end{table*}

To add to the puzzle of $\omega$ Cen, it has also been linked to tidal streams in the MW halo.  \citet[][hereafter \citetalias{2019NatAs.tmp..258I}]{2019NatAs.tmp..258I} used {\it Gaia} DR2 to uncover a stellar stream that they named \textit{Fimbulthul}, which they associated to $\omega$ Cen on the basis of its similar orbit, distance, and stellar populations, even although it is located almost 28$^\circ$ away. This comparison is based on five \textit{Fimbulthul} stream stars and a matched filter analysis, with additional support from an N-body simulation performed by the authors. Since then, other studies have searched for additional \textit{Fimbulthul} members, and/or other streams that could be attributed to $\omega$ Cen.  For example, \citet[][]{2024ApJ...967...89I} increase the number of \textit{Fimbulthul} member stars to 22, and show that another stream, which they call {\it stream \#55}, resides in the same location in action-energy space. They argue both streams are trailing debris from $\omega$ Cen, which is again broadly consistent with the predictions of their N-body model. On the other hand, the connection between \textit{Fimbulthul} and $\omega$ Cen is less clear when considering chemistry. \citet[][]{2020MNRAS.491.3374S} searched for \textit{Fimbulthul} members in the GALAH dataset \citep[][]{2015MNRAS.449.2604D,2018MNRAS.478.4513B}, and identified two stars in the stream with similarly enhanced s-process elements to the main body of $\omega$ Cen, suggesting the first chemical connection of \textit{Fimbulthul} to the cluster. Also using GALAH, \citet[][]{2023MNRAS.524.2630Y} blindly searched for $\omega$ Cen stars in the MW halo by applying an unsupervised machine-learning technique. Although they found several stars with light element abundances consistent with $\omega$ Cen scattered throughout the halo, none were spatially coincident with \textit{Fimbulthul}. Finally, an analysis of r-process abundances in the stream conducted by \citet[][]{2021ApJ...912...52G} led the authors to suggest that it is more likely that \textit{Fimbulthul} and $\omega$ Cen share a progenitor, not that \textit{{Fimbulthul}} is debris from $\omega$ Cen itself. These conflicting interpretations call for more exploration to fully understand the connection between \textit{Fimbulthul} and $\omega$ Cen. 

A recent simulation of the disruption of $\omega$ Cen by \citet[][]{2026ApJ..1000..176Z} indicates that its tidal tails should maintain the same stellar population mix as that seen within the central regions, with no expected population gradient. This is in agreement with the findings of \citet[][]{2025MNRAS.537.2752K}, who saw the same multiple stellar population signatures in the main body and outer regions of $\omega$ Cen, including its tidal tails. This analysis, which was performed with synthetic photometry from Pristine DR1 \citep[][]{2024A&A...692A.115M} and {\it Gaia} astrometry, underscores the importance of considering both metallicities {\it and} kinematics for studying tidal debris around GCs. However, the relatively shallow Pristine synthetic photometry meant that only the sparse red giant branch (RGB) component of the tails could be explored. 

In this study, we present the results of a spectroscopic survey of the peripheral regions and tidal tails of $\omega$ Cen, following up on the identification of probable members in these parts taken from  \citetalias{2021MNRAS.507.1127K}. The only previous spectroscopic study to specifically target the outer regions of the cluster is that of \citet[][hereafter \citetalias{2012ApJ...751....6D}]{2012ApJ...751....6D}, who covered cluster-centric radii of 25 to 45 arcmin and found a flat line-of-sight velocity profile, tentatively attributed to the dynamical evolution of the cluster within the Galactic gravitational field. Our study, which probes more than four times further out, also nicely complements the recent oMEGACat series of papers that aims to improve understanding of the main body of $\omega$ Cen \citep[e.g., ][]{2023ApJ...958....8N,2024ApJ...970..192H,2024ApJ...970..152N}. In the subsequent sections, we will describe the methods used to reduce and analyse the spectra (Sec. 2), identify $\omega$ Cen stars and explore their kinematics and metallicity as a function of position (Sec. 3), and discuss our findings (Sec. 4).

\begin{figure*}
 \begin{center}  
    \includegraphics[width=\textwidth]{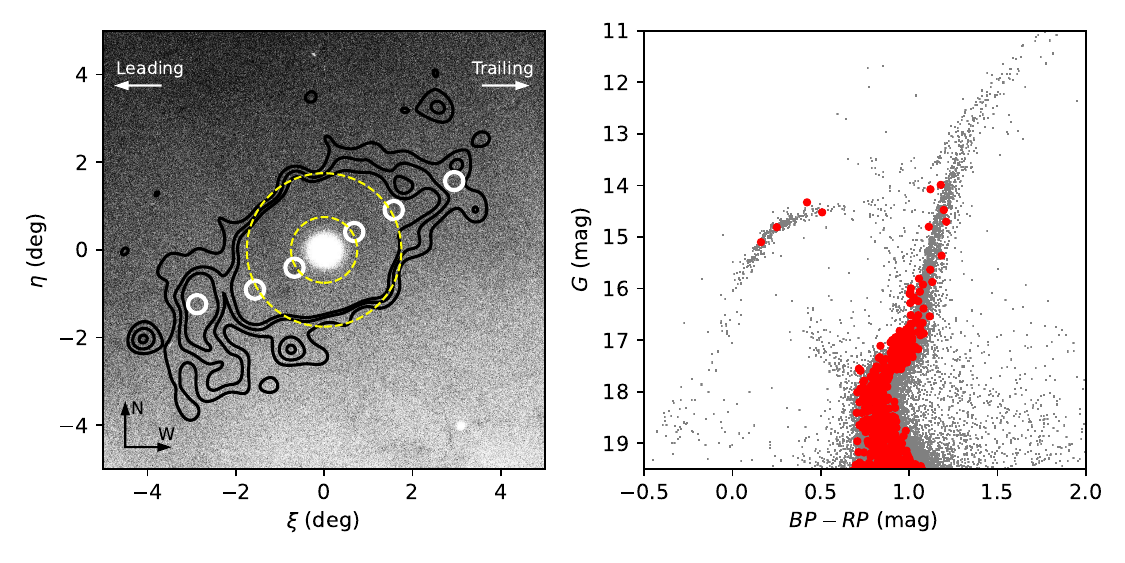}
  \end{center}
\caption{Left: Locations of our six fields (white circles) overlayed on a optical image cutout of $\omega$ Cen from DR3 \citep[][]{2023A&A...674A...1G}. Additionally, we have included the contours of the tidal tails from \citetalias{2021MNRAS.507.1127K}, and the tidal (Jacobi) radii through inner (outer) yellow dashed circles \citep[46.4 and 106.9 arcmin respectively][]{2018MNRAS.474.2479B,2019MNRAS.485.4906D} We indicate the direction of the trailing and leading tails through labeled arrows at the top of the image. Right: The CMD of our targets (red) overlayed on that of all stars lying within a cluster-centric radius of 45 arcmin. The photometric system is DR3: that is, BP, RP and G magnitudes.}
\label{fig:fig_1_position}
\end{figure*}

\section{The Data}
\subsection{Observations}\label{sec:2_1_Observations}
The observations utilised in this study were obtained with the Fibre Large Array Multi-Element Spectrograph (FLAMES). FLAMES is a multi-object spectrograph mounted on UT2 of the Very Large Telescope (VLT), located on Cerro Paranal, Chile \cite[][]{2002Msngr.110....1P}. FLAMES is fed by two different instruments, but, in our instance, we utilised only the fibre-fed spectrograph, GIRAFFE. GIRAFFE allows for up to 130 fibres to be placed for science, sky and dark observations, with spectral resolutions ranging from medium to high. In this instance, we used the LR8 filter, which covered wavelengths of 8200-9400 \AA\ with a resolution of R=6500. Importantly, this spectral range covers the \ion{Ca}{II} triplet, which is used for line-of-sight velocity and [Fe/H] measurements. We measured stars in six fields, each of which was observed with $2 \times 1300$s exposures, during February and March 2022. Table \ref{tab:obs} summarises the observations.  

The main goal of the study was to confirm and characterise stars in the periphery and tidal tails of $\omega$ Cen, based on the membership probabilities presented in \citetalias{2021MNRAS.507.1127K}. Our innermost field centers were selected to lie at cluster-centric radii of 47 arcmin\footnote{We adopt ($\alpha,\delta) = (201.697, -47.469$) deg for the centre of $\omega$ Cen, taken from \citet{2021MNRAS.505.5978V}.} so as to have some overlap with \citetalias{2012ApJ...751....6D}, while our outer fields sampled the leading and trailing tidal tails at cluster-centric radii of 1.8 deg and 3.2 deg (see left panel of Fig \ref{fig:fig_1_position}).  Within each field, we used the membership probabilities ($P_\text{mem}$) derived in \citetalias{2021MNRAS.507.1127K} to allocate fibres; highest priority was assigned to stars with $P_\text{mem}=1$, with priority decreasing progressively for lower $P_\text{mem}$ values. Most targets are located on the lower RGB or at the main sequence turn-off (MSTO). There are also a small number of blue horizontal branch (BHB) candidates, while the outermost fields also contain targets on the main sequence. The CMD of the targets is presented in the right panel of Fig \ref{fig:fig_1_position}. Lastly, we have re-observed three stars from \citetalias{2012ApJ...751....6D} in order to validate our measured line-of-sight velocities against theirs. In total, we have obtained spectra for 616 stars.

Since the observations were obtained, the Bayesian inference technique of \citetalias{2021MNRAS.507.1127K} has undergone various improvements. Specifically, the spatial model has been adjusted to better capture the behaviour at the boundary between the GC main body and the extended structure, and additional structural parameters are now fit.  A detailed explanation of the latest model is presented in a forthcoming paper (Kuzma et al., in prep). These improvements result in small changes to the membership probabilities -- for example, the number of high-probability stars, defined as $P_\text{mem}>0.3$, changes from 169 \citepalias{2021MNRAS.507.1127K} to 179 (Kuzma et al., in prep). In order to be consistent with our latest analysis, we will use these updated membership probabilities for the rest of the paper. We have verified that this has no bearing on our results. 

\subsection{Data Reduction}
The spectra were reduced using the pipelines available from ESO, \textsc{EsoReflex}\footnote{\url{https://www.eso.org/sci/software/esoreflex/}} \citep[][]{2013A&A...559A..96F}, and we used the default calibration set of the associated GIRAFFE workflow. This workflow follows the standard steps of spectroscopic reduction: bias-subtraction, flat-fielding, wavelength calibration and heliocentric velocity correction. Additionally, we utilised the additional sky-subtraction and cosmic ray subtraction using \textsc{PyCosmic} \citep[][]{2012A&A...545A.137H}, which is available as an extension package with \textsc{EsoReflex}. 

The line-of-sight heliocentric velocities, $V_\text{h}$, and stellar metallicities, [Fe/H], were calculated using the freely-available automated spectroscopic pipeline \textsc{rvspecfit} \citep[][]{2011ApJ...736..146K,2019ascl.soft07013K}. The pipeline can fit multiple spectra of the same object simultaneously, allowing for precision velocities and atmospheric parameters to be calculated using a maximum-likelihood technique. For this work, we supplied either both the individual (uncombined) sky-subtracted spectra for each star or just the single spectra if the star was only observed once (e.g., due to fiber unavailability between plates) and recorded the measurements and their associated uncertainties provided by \textsc{rvspecfit}. We also compared the velocity output of \textsc{rvspecfit} to the velocities of the spectral template cross-correlation technique used in \citet[][]{2022MNRAS.512..315K}, and found a near one-to-one correspondence. 

Targets where either \textsc{rvspecfit} was unable to perform the fit effectively, or where the $V_\text{h}$ uncertainties were greater than 5 km s$^{-1}$ (typically due to poor signal-to-noise) were removed.   This affected 23 stars, including 13 stars with $P_\text{mem}>0.3$. This corresponds to 4 per cent of the original sample, and resulted in a final sample of 593 stars with which we search for $\omega$ Cen members. Of these, 166 stars have a high-probability (i.e, $P_\text{mem}>0.3$) of membership from our Bayesian technique. As noted previously, three stars are in common with \citetalias{2012ApJ...751....6D}, which gave us an opportunity to search for any systematic velocity offset in our measurements. We measure $(\overline{V_\text{h}-V_\text{\citetalias{2012ApJ...751....6D}}})=0.7+/-0.5$ km s$^{-1}$, which places our velocities in excellent agreement, and well within the measurement uncertainties of our data. Indeed, the median radial velocity uncertainty of our final sample is $\sigma_{V_\text{h}}=1.5$ km s$^{-1}$.

\subsection{Coordinate transformations}\label{sec:coord_trans}
We identify stars in the periphery and tidal extensions of $\omega$ Cen by combining existing membership probabilities with kinematic and chemical information from spectroscopy. Due to the large angular extent of $\omega$ Cen, the 3D kinematics have the potential to be severely affected by projection effects, such as perspective rotation and curvature in the plane of the sky \citep[e.g.,][]{2006A&A...445..513V}. Without correcting for such effects, there will be an apparent rotation in the velocities that can affect the way in which we assign membership \citep[][]{1961MNRAS.122....1F}. To account for this, we performed coordinate transformations and velocity corrections. First, we transformed the sky coordinates ($\alpha$,$\delta$) and proper motions ($\mu^*_{\alpha}$,$\mu_{\delta}$)\footnote{$\mu^*_{\alpha} \equiv\mu_{\alpha}\cos{\delta}$} onto a tangential gnomonic coordinate system ($\xi,\eta$) with the associated proper motion ($\mu_{\xi}$,$\mu_{\eta}$), where north and west are in the positive directions. This was done using equation 2 in \citet{2018A&A...616A..12G}, and we propagated the uncertainties and co-variances appropriately through the transformation. Additionally, we corrected our radial velocities for perspective motion. This was done using the relationships from eq. 6 in \citet[][]{2006A&A...445..513V}:

\begin{equation}
\Delta V_\text{h} = 1.379 \times 10^{-3}(\xi \mu_\xi + \eta \mu_\eta)\ D \ \textrm{km s}^{-1}\label{eq:rot_cor}
\end{equation}

and we subtracted the $\Delta$ $V_\text{h}$ component from the measured heliocentric velocity. In eq. \ref{eq:rot_cor},  $\mu$ is the corresponding bulk proper motion of $\omega$ Cen ($\mu_\xi,\mu_\eta$) = ($-3.25 \pm 0.01$, $-6.75 + 0.01$) (Kuzma et al., in prep), which is consistent with the proper motion reported by \citet[][]{2021MNRAS.505.5978V}, and $D$ is the heliocentric distance of $\omega$ Cen.  We will use these corrected velocities for the rest of this paper. For consistency, we measured the dynamical heliocentric distance to $\omega$ Cen using the 1D perspective motion correction for the proper motion in the radial direction. This involved selecting stars with a revised membership probability of $\geq 0.9$ that lie within the nominal tidal radius of 46.4 arcmin \citep[][]{2019MNRAS.485.4906D}. Using eq. 4 from \citet[][]{2018MNRAS.481.2125B}, and setting the bulk line of sight velocity to $232.78\pm 0.21$km s$^{-1}$ from \citet[][]{2021MNRAS.505.5978V}, we fit for the distance. We find a dynamical distance of 5.29 $\pm$ 0.2 kpc, in firm agreement with distance measurements derived through other methods \citep[e.g.,][]{2021ApJ...908L...5S,2021MNRAS.505.5978V,2025ApJ...983...95H}. Additionally, we created a new coordinate system denoted ($x''$,$y''$), which is the coordinate frame $(\xi, \eta)$ rotated by the stream position angle \citetalias[$\theta$ = 124 deg;][]{2021MNRAS.507.1127K} so that the major axis of the debris is aligned with the $x''$-axis \citep[see also][]{2006A&A...445..513V}.

\begin{figure*}
 \begin{center}  
    \includegraphics[width=\textwidth]{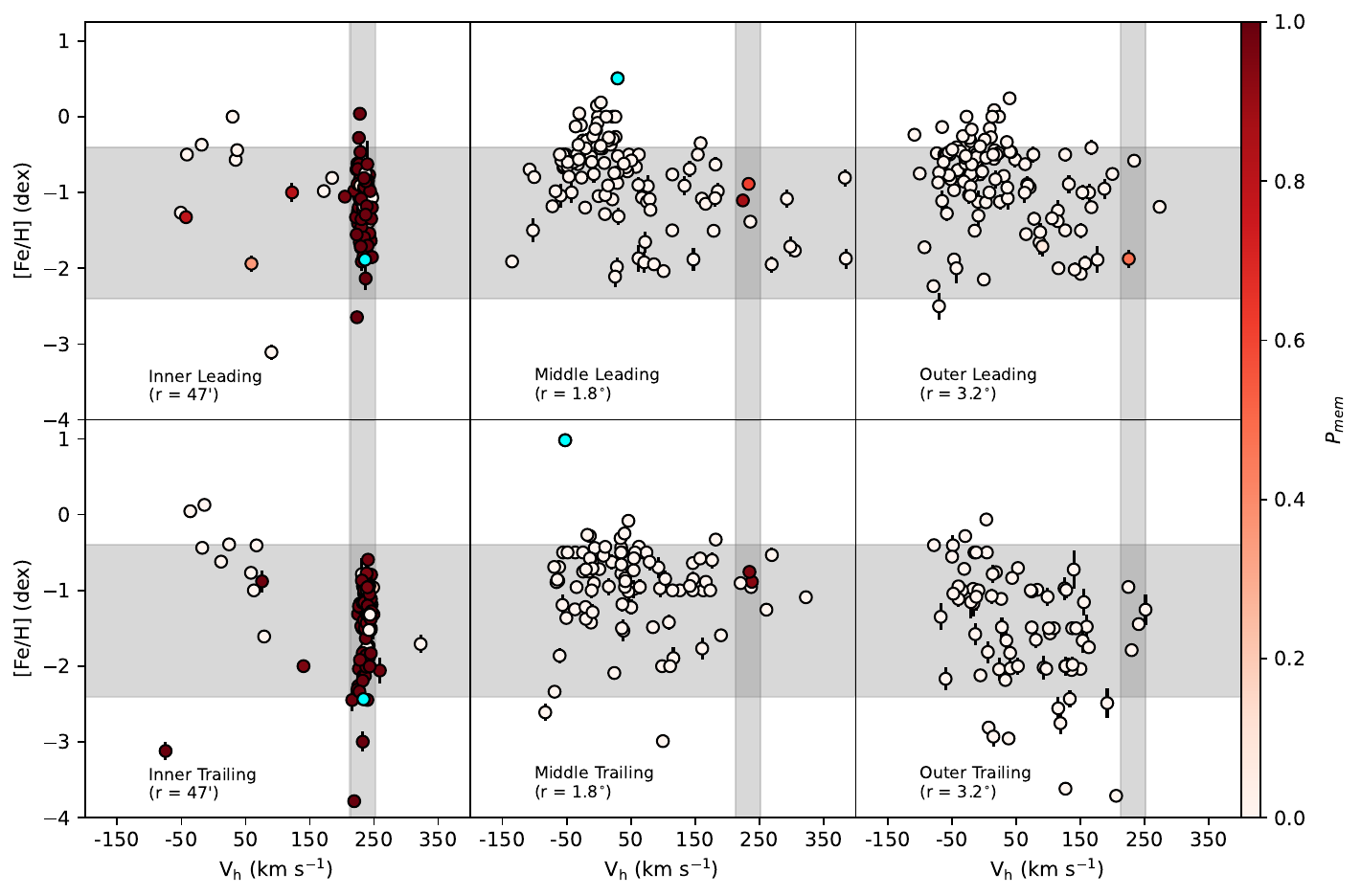}
  \end{center}
\caption{Diagnostic plots showing [Fe/H] against $\textrm{V}_\textrm{h}$ for each observed field along the stream, denoted in the lower left of each panel. Each target has been colour coded by their respective probabilities, with white corresponding to the lowest membership probability, and dark red corresponding to the  highest. Additionally, the BHB stars selected for observation are shown in cyan. The vertical shaded region indicates the known line-of-sight velocity of $\omega$ Cen \citep[232 km s$^{-1}$,][]{2021MNRAS.505.5978V}, and is of width three times 6.7 km s$^{-1}$ either side, the velocity dispersion at the tidal radius reported by \citetalias{2012ApJ...751....6D}. The shaded horizontal region indicates the mean metallicities of the most metal-rich ($-0.5$ dex) and metal-poor ($-2.4$ dex) populations of $\omega$ Cen. The panels show fields of increasing cluster-centric radius from left to right.}
\label{fig:fig_x_RV_FEH}
\end{figure*}

\section{Results and Discussion}

\subsection{Identifying Stars in the Periphery and Tidal Tails of $\omega$ Cen}
The addition of spectroscopy allows us to assess which stars flagged as high-probability members of $\omega$ Cen in \citetalias{2021MNRAS.507.1127K}, which relies on astrometry and photometry alone, are also consistent in terms of their $V_\text{h}$ and [Fe/H]. Fig. \ref{fig:fig_x_RV_FEH} shows our measurements of $V_\text{h}$ and [Fe/H], with fields sampling the leading (trailing) arm along to top (bottom) row, increasing in cluster-centric radius from left to right.
The shaded regions indicate where we expect $\omega$ Cen stars to lie -- these are defined by the main body metallicity spread of $-2.4$ to $-0.5$ dex, and the line-of-sight velocity range 212--252 km s$^{-1}$. The latter is centered on $\omega$ Cen's radial velocity \citep[232 km s$^{-1}$,][]{2021MNRAS.505.5978V}, and encompasses three times the line-of-sight velocity dispersion of 6.7 $\textrm{km s}^{-1}$, measured at the tidal radius  \citepalias{2012ApJ...751....6D}.  

Encouragingly, a significant number of high-probability stars fall within the shaded velocity region, indicating that they are likely to be genuine members.  In the inner fields, the majority of the target stars are moving with the same line-of-sight velocity as $\omega$ Cen. We also identify two BHB stars in the inner fields that are co-moving with the GC. In the middle to outer fields, the target density of high-probability stars is much lower but we still find a small number of stars with radial velocities that are consistent with membership. In total, out of 166 stars with $P_\text{mem}>0.3$, we find 160 stars to have $\textrm{V}_\textrm{h}$ within the expected line-of-sight velocity range of $\omega$ Cen, plus two BHB candidates.

A similar picture can be seen when we compare the [Fe/H] of our targets with the known metallicity distributions of $\omega$ Cen. When considering [Fe/H] determination, the precision of the measurement varies across different evolutionary phases. We find RGB stars  to [Fe/H] uncertainties of 0.02 dex, while the uncertainties increase to approximately 0.10 dex for MSTO and MS stars and 0.04 dex for BHB stars. To account for these varying precisions, a star is considered consistent with $\omega$ Cen if the interval defined by $3\sigma_{\rm [Fe/H]}$ around its measured metallicity overlaps the cluster metallicity range of $-2.4 < \mathrm{[Fe/H]} < -0.5$ dex. Of the stars with $P_\text{mem} > 0.3$, the majority lie within the known metallicity spread, with only a small number of stars outside this range. For the 166 $P_\text{mem}>0.3$ stars, only six stars have metallicities severely offset from the known populations of $\omega$ Cen (two with [Fe/H] $> -0.3$ dex, and four with [Fe/H] $< -2.5$ dex), leaving 160 stars. Again, the two BHB stars in the inner field have consistent metallicities with the main population.

\begin{figure*}
 \begin{center}  
    \includegraphics[width=\textwidth]{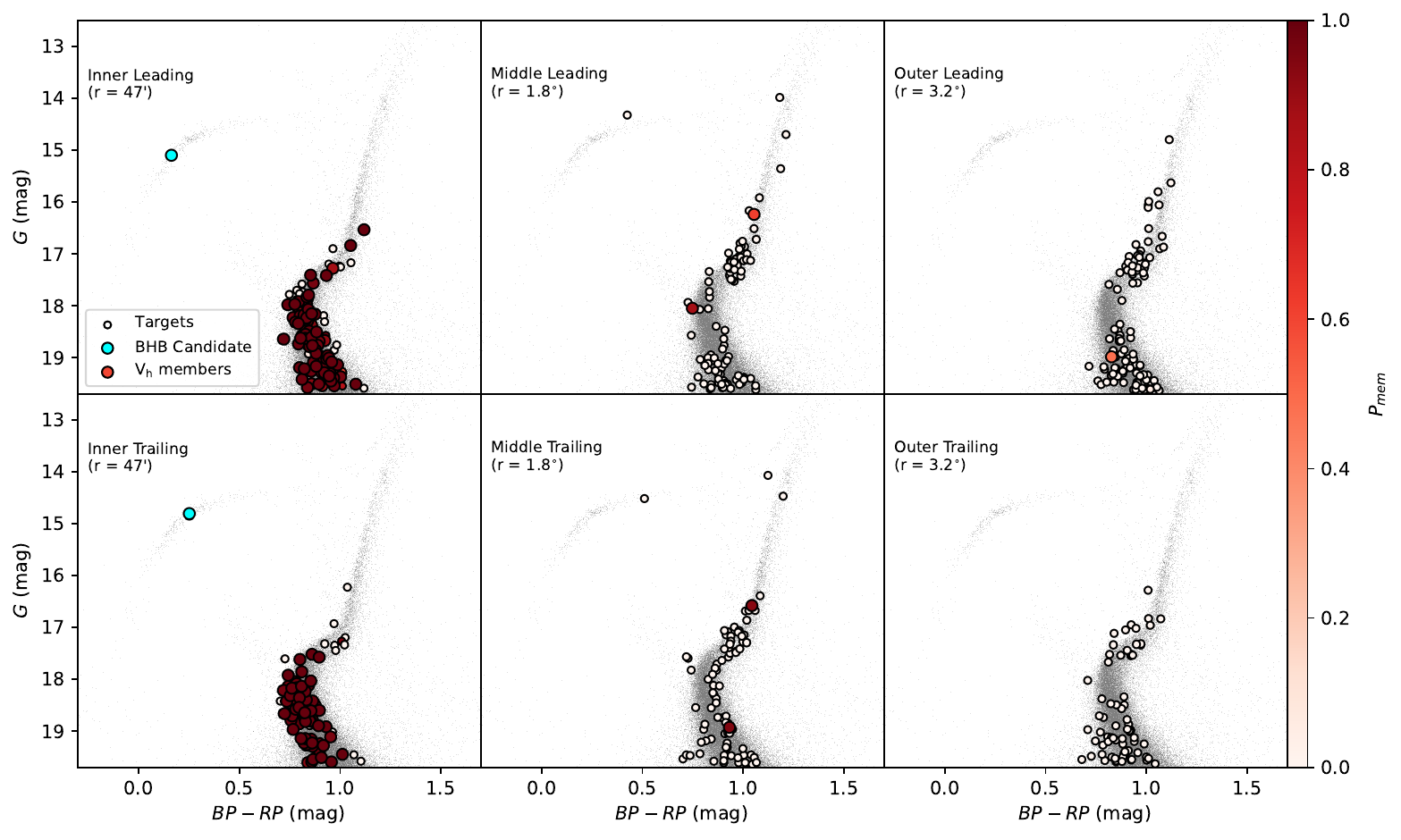}
  \end{center}
\caption{Diagnostic plots demonstrating the CMD for each observed field along the stream, denoted in the upper right of each figure. The large points indicate the 155 $P_{\textrm{mem}} > 0.3$ targets we identify as potential $\omega$ Cen members and are coloured according to their $\textrm{P}_\textrm{mem}$ value. Cyan points are the confirmed BHB candidates, and the white points are the rest of the observed stars. In each figure, we have the CMD of $\omega$ Cen from the right plot of Fig. \ref{fig:fig_1_position}.}
\label{fig:fig_x_CMD}
\end{figure*}

In summary, considering stars that lie within the expected bounds of both $V_\text{h}$ and [Fe/H], and excluding the ten stars that are outside those bounds, we find 155 out of 166 (93\%) of the high-probability \citetalias{2021MNRAS.507.1127K} stars are consistent with $\omega$ Cen membership. Five of these are located beyond the Jacobi radius \citep[106.9 arcmin,][]{2018MNRAS.474.2479B}.  While the two innermost BHB stars are likely $\omega$ Cen members, the BHB stars in the middle fields are excluded due to their very different line-of-sight velocities. Overall, we nominate 157 stars as members of the outskirts of $\omega$ Cen and its tidal tails (see Table \ref{tab:target_flow} for a summary of this selection process). The locations of these stars (outlined by the larger symbols) are overlaid on the $\omega$ Cen CMD in Fig. \ref{fig:fig_x_CMD}. We see that these sample all stages of the CMD - main sequence, MSTO, RGB and BHB. We provide the full list of observed stars with their corresponding membership probabilities in Table \ref{tab:target_list}, noting those that we nominate as {\it bona fide} $\omega$ Cen members.  It is worth noting that while the \citetalias{2021MNRAS.507.1127K} technique includes a model for the MW field, there may still be residual field contamination in the final sample. We explored the
predictions of the Besançon MW models \citep[][]{2003A&A...409..523R} to assess this
potential contamination.  In a field of view the same size as VLT/FLAMES, we
randomly selected the same number of stars as those observed in
our observations across the fields (130 stars), and repeated this 1000 times.
Considering the photometric limit of Gaia (G = 20 mag), and the kinematic (line-of-sight velocities and proper motions) and metallicity cuts applied to select $\omega$ Cen stars (see above), we found
that less than 1 per cent of the realisations returned one or more stars.  Therefore, we are confident that MW field contamination is negligible, despite the small number of stars identified in the middle and outer fields.

\begin{table}
\caption{Construction of Final Sample.}
\centering
\begin{tabular}{lc}
\hline
Stage & Target numbers\\
\hline\hline
Initial stellar sample & 616\\
Stars remaining after quality cut & 593 \\
Stars with $P_{\textrm{mem}} >=0.3$ &  166 \\
$P_{\textrm{mem}} >=0.3$ stars consistent with $\omega$ Cen & 155\\
No of HB stars & 2\\
Final sample& 157\\
\hline \hline
\label{tab:target_flow}
\end{tabular}
\end{table}

\begin{table*}
\caption{List of observed targets. Our targets take the naming convention of *\_T\_* where the first asterisk indicates which field it is located in: preceding O, M, and I relates to the outer, middle and inner field, and the proceeding letter relates to whether it is found in the trailing (T) or leading (L) tail.  The later asterisk relates to the star number in that field. The columns are: 1) Star ID, 2) DR3 Source ID, 3) Right ascension (J2000), 4) Declination  (J2000), 5) DR3 G-band magnitude, 6) BP-RP colour index, 7) Heliocentric velocity, and 8) uncertainty, 9) [Fe/H], and  10) uncertainty, and 11) membership probability from Kuzma et al. (in prep). {\it This table is fully available as online material}.}
\centering
\begin{tabular}{lrrrrrrrrrc}
\hline
Star ID & DR3 Source ID & R.A. & Dec. & $G$ & $V_\text{h}$ & $\sigma_{V_\text{h}}$ & $\textrm{[Fe/H]}$ & $\sigma_{\text{[Fe/H]}}$ & $P_\text{mem}$ & Member\\
&&(J2000, deg)&(J2000, deg)&(mag)&(mag)& (km s$^{-1}$)&(km s$^{-1}$)&(dex)&(dex)&\\
\hline\hline
OT\_T\_325 & 6086552125451670000 & 197.174 & --45.862 & 16.83 & --49.93 & 0.39 & --0.56 & 0.03 & 0.00 & N\\
OT\_T\_363 & 6086559164897950000 & 197.178 & --45.786 & 19.19 & 251.28 & 3.68 & --1.25 & 0.20 & 0.00 & N\\
OT\_T\_358 & 6086558065386160000 & 197.19 & --45.838 & 18.97 & 11.69 & 1.61 & --1.07 & 0.10 & 0.00 & N\\
OT\_T\_364 & 6086559336696750000 & 197.236 & --45.744 & 19.00 & 126.38 & 1.87 & -2.00 & 0.08 & 0.00 & N\\
OT\_T\_362 & 6086559061818780000 & 197.24 & --45.768 & 17.77 & 74.08 & 0.80 & --1.49 & 0.05 & 0.00 & N\\
\hline \hline
\label{tab:target_list}
\end{tabular}
\end{table*}

\begin{figure}
 \begin{center}  
    \includegraphics[width=\columnwidth]{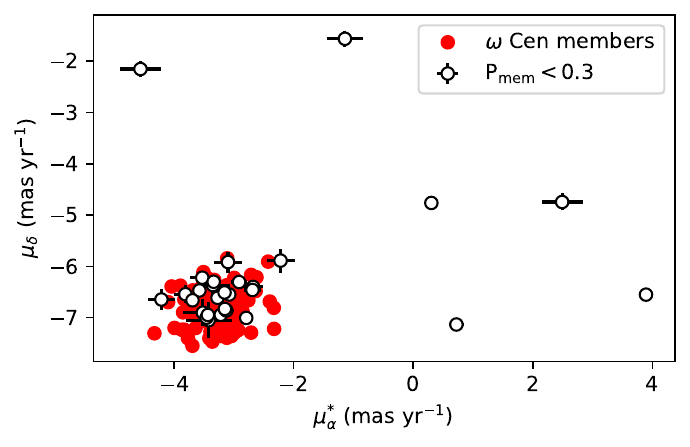}
  \end{center}
\caption{Proper motion distribution of the $\textrm{P}_\textrm{mem} \leq0.3$ stars (white) that have consistent [Fe/H] and line-of-sight velocities with the nominated $\omega$ Cen members, which have $\textrm{P}_\textrm{mem} >0.3$ (red). 23 of the former stars are consistent with the nominated members.}
\label{fig:fig_low_pm}
\end{figure}

 Upon closer inspection of the low-probability ($P_\textrm{mem}\leq0.3$) stars in Fig. \ref{fig:fig_x_RV_FEH}, we find a number (29) that have line-of-sight velocity and [Fe/H] consistent with $\omega$ Cen membership. Several of these stars could plausibly be members of $\omega$ Cen and its tidal tails. Fig. \ref{fig:fig_low_pm} shows the $\it Gaia$ proper motions of these stars alongside those of our nominated members.  We find that 23 of these stars lie within the region expected for $\omega$ Cen, and hence move coherently with it. The majority of these stars are located in the inner field near the tidal radius, with only three stars located in the middle and outer fields. The reason why these stars have not been assigned higher membership probabilities could be due to some shortcomings of our modelling technique, or the fact that many of these stars are faint (G > 18 mag) so may have less accurate astrometric solutions.  Revisiting these stars with  improved {\it Gaia} DR4 astrometry could allow us to reclassify them, but the relatively small number of stars affected (roughly 4 per cent of the total targets, or 14 per cent of our nominated members) indicates that our methodology is largely robust. 

\begin{figure*}
 \begin{center}  \includegraphics[width=\textwidth]{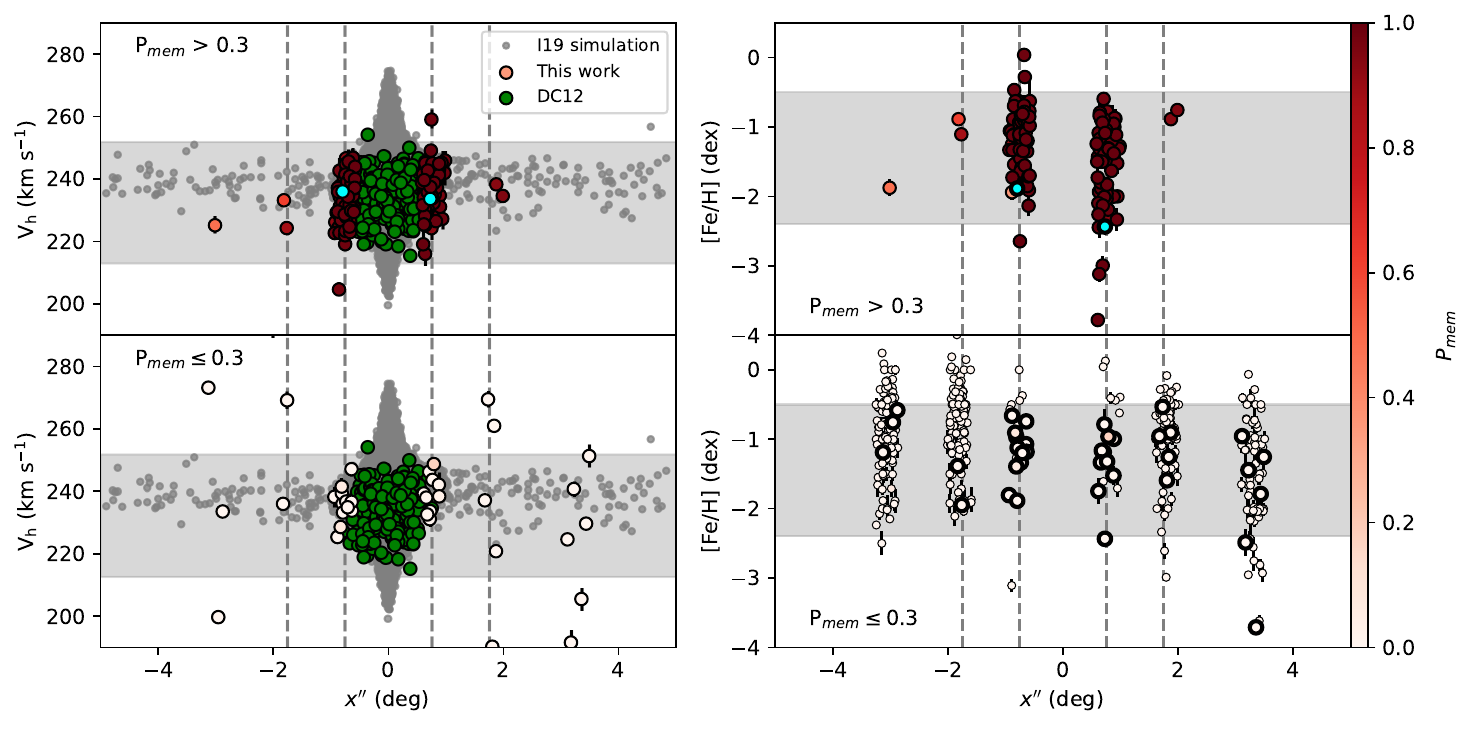}
  \end{center}
\caption{$\textrm{V}_\textrm{h}$ and [Fe/H] as a function of position along the direction of the $\omega$ Cen tidal tails, denoted by $x''$. The horizontal shaded regions in both panels are the same as those in Fig. \ref{fig:fig_x_RV_FEH}.  Left: $\textrm{V}_\textrm{h}$ of stars, colour-coded by $P_\text{mem}$ and split between $P_\text{mem}>0.3$ in the top panel, and $P_\text{mem}<=0.3$ in the bottom panel. Overplotted are the velocities from the simulation of $\omega$ Cen's disruption by \citetalias{2019NatAs.tmp..258I} in gray, and the velocities of \citetalias{2012ApJ...751....6D} in green. The vertical dashed lines in all figures correspond to the tidal (inner two lines) and Jacobi (outer two lines) radii. Right: Same as the left but for [Fe/H], and without the simulated system of \citetalias{2019NatAs.tmp..258I}. The lower plot highlights low-probability stars which lie within the velocity range displayed in the left panel.}
\label{fig:fig_x_RV}
\end{figure*}

\subsection{Radial Trends in Kinematics and Metallicity}\label{sec:stream_dist}

Although the line-of-sight velocity and the metallicity distribution of $\omega$ Cen are well known, exactly how these properties vary into the periphery and along the tidal tails is not. In Fig. \ref{fig:fig_x_RV}, we examine $\textrm{V}_\textrm{h}$ and [Fe/H] as a function of radial position along the stream, where we now use the rotated coordinate frame which places the stream along the $x''$ axis (see Sec. \ref{sec:coord_trans}). The left figure shows $\textrm{V}_\textrm{h}$, while the right figure shows [Fe/H]. We split them into two groups,  $P_\text{mem} > 0.3$ (top panel) and $P_\text{mem} \leq 0.3$ (bottom panel), to show how the high- and low-probability stars compare to each other.  The shaded regions are the same as those shown in Fig. \ref{fig:fig_x_RV_FEH} 

For comparison, we also show the predictions of the N-body simulation of 
$\omega$ Cen's disruption that was presented in \citetalias{2019NatAs.tmp..258I}. An important aspect of this simulation was the inclusion of rotation in the progenitor, without which the authors failed to match the morphology of the {\it Fimbulthul} stream.  The \citetalias{2019NatAs.tmp..258I} model is based on {\it Gaia} DR2 kinematics and is underpinned by the family of rotating models presented in \citet[][]{2012A&A...540A..94V}.  Although \citetalias{2019NatAs.tmp..258I} caution that they have taken a simplified approach to modelling the disrupting GC, this simulation still presents a useful benchmark for comparison.  In particular, the included rotation could potentially have an effect on the kinematics of outer regions of $\omega$ Cen, and the tidal tails \citep[e.g.,][]{2013ApJ...772...67B}.   We performed the same coordinate transformations to the simulated dataset while also correcting for perspective motion, and show the results in grey points in the left side of Fig. \ref{fig:fig_x_RV}. We also add the similarly-transformed stars from \citetalias{2012ApJ...751....6D} as these lie inside the tidal radius of the cluster, where we have no coverage.  Note that in this representation, $x''$ does not differentiate between stars with different $y''$ near $x'' = 0$. Both this work and \citetalias{2012ApJ...751....6D} do not measure stars in the central regions of $\omega$ Cen, which is where the largest velocity dispersion is expected.

In the top-left figure, we see good consistency between the line-of-sight velocities of our $P_\text{mem}>0.3$ stars and the simulation predictions.   Furthermore, our spread in velocities matches well with that seen by  \citetalias{2012ApJ...751....6D} at smaller radii.  The lower-left figure demonstrates that, as noted previously, the inner regions (and to a lesser extent the other fields) have several stars with lower membership probabilities that also possess line-of-sight velocities consistent with the stream. The simulation predicts that the line-of-sight velocity gradient is flat along the stream, at least within the radial range probed by our observations. Although we have few nominated members beyond the main body of $\omega$ Cen, their velocities are generally consistent with this trend. At $x'' = 3$ deg, our outermost $P_\text{mem} \ge 0.3$ star is offset by $\sim$ 7 km s$^{-1}$ from the GC bulk velocity; further measurements are needed in order to test the significance of this. 

The right panels of Fig. \ref{fig:fig_x_RV} show how metallicity varies as a function of $x''$. The inner fields follow the known populations of $\omega$ Cen, except for a the small number of stars outside the metallicity range of --2.4 to --0.5 dex \citep[][]{2020AJ....159..254J}. The stars in the two middle fields in the top right panel cluster around $-1.0$ dex on both sides, consistent with one of the more metal-rich populations of $\omega$ Cen found by \citet[][]{2014ApJ...791..107V}. The singular star in the outermost (leading tail) field is more metal poor at $-2.0$ dex, but we cannot make any conclusions about the metallicity of the populations at these distances, even if we consider how the lower probability stars with $\textrm{V}_\textrm{h}$ consistent with the stream (the bold points in lower right figure).

Lastly, we consider how our measured metallicities of stars in the $\omega$ Cen periphery and tidal tails compare to those of stars in the \textit{Fimbuthul} stream. There are only a handful \textit{Fimbuthul} stars with reported metallicities in the literature. \citetalias{2019NatAs.tmp..258I} present [Fe/H] measurements for four stars that range between $-1.4$ to $-1.8$ dex, with an associated mean uncertainty of 0.11 dex. Additionally, \citet[][]{2020MNRAS.491.3374S} identify three stars as \textit{Fimbulthul} members, and these stars have [Fe/H] ranging from $-1.5$ to $-1.9$ dex, with an average uncertainty of 0.08 dex.  \citet[][]{2020MNRAS.491.3374S} further identify enhanced aluminium and sodium in their candidates, which forms a stronger basis for their candidature as former members of $\omega$ Cen.  The five members that we detect in the tidal tails have metallicities of $-1.0$ to $-2.0$ dex and hence are broadly consistent with those in the \textit{Fimbulthul} stream. However, unlike \citet[][]{2020MNRAS.491.3374S}, we lack measurements of the light and heavy element abundances required to definitively establish an association between \textit{Fimbulthul} and the candidate stars identified in the tidal tails of $\omega$ Cen. Nevertheless, these candidates offer promising targets for future spectroscopic follow-up, enabling a more conclusive test of their origin.

\subsection{Comparison to Pristine}
\begin{figure}
 \begin{center}  
    \includegraphics[width=\columnwidth]{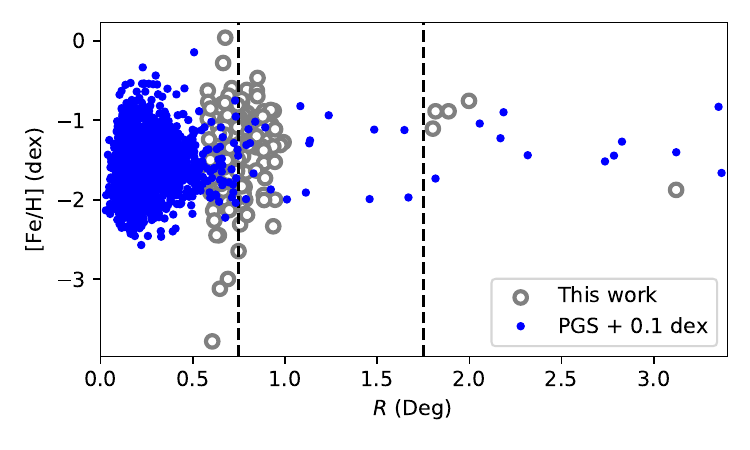}
  \end{center}
\caption{Metallicity distribution of our $P_{mem}>0.3$ stars as a function of radius, with the PGS sample from \citet[][]{2025MNRAS.537.2752K} overlaid. The vertical dashed lines indicate the tidal and Jacobi radii of $\omega$ Cen.}
\label{fig:PGS_R}
\end{figure}

The first data release (DR1) of the Pristine photometric survey provided photometric metallicities based on CaHK-band photometry \citep[][]{2024A&A...692A.115M}. An associated product of the Pristine DR1 is the Pristine-Gaia-Synthetic (PGS) catalogue, which provides photometric metallicities over the whole sky based on synthetic CaHK-photometry calculated through {\it Gaia} XP spectra. \citet[][]{2025MNRAS.537.2752K} used the PGS catalogue to identify the stars in the outer regions of $\omega$ Cen and into the tidal tails. To compare the stars studied in this work to the PGS sample, we first checked the calibration of the different metallicity scales. Due to the deep nature of the present study, we have only ten stars in common with the sample of \citet[][]{2025MNRAS.537.2752K}. We find a mean-weighted [Fe/H] difference of $(\text{[Fe/H]}-\text{[Fe/H]}_{PGS}) = 0.2 \pm 0.2$ dex, which reduces to $0.1 \pm 0.1$ dex when considering the nine stars in the metallicity regime of $\text{[Fe/H]} < -1.0$ dex, the regime where Pristine is considered to be most accurate \citep[see section 7.4.3 in][]{2024A&A...692A.115M}.  Fig. \ref{fig:PGS_R} compares the metallcities of the high-probability stars from \citet[][]{2025MNRAS.537.2752K} to our $P_{mem}>0.3$ stars as a function of radius, considering the 0.1 dex offset described above. At the tidal radius, where our spectroscopic observations begin (the left vertical dashed line), we see a much broader [Fe/H] distribution than what was seen in the PGS dataset. While the datasets show good overall agreement in [Fe/H], the difference in the [Fe/H] dispersion is striking. We suspect that this could be due to the fact that magnitude range of the PGS has very limited overlap with our spectroscopic sample, therefore we may merely be seeing a sampling issue with considerably fewer stars available in the PGS at this radius and beyond. There may be further effects due to the PGS metallicities being underestimated at high metallicity,  and overestimated at low ones within crowded regions (see Fig. 9 in \citealt[]{2025MNRAS.537.2752K}).

\section{Conclusion}
In this paper, we present VLT/FLAMES spectroscopic observations of candidate members in the peripheral regions and tidal tails of the massive GC, $\omega$ Cen. We sampled six locations spanning 6 degrees, ranging from a cluster-centric radii of 45 arcmin to 3 degrees (approximately 75 to 300 pc), extending more than four times further out than any previous study of the $\omega$ Cen main body. Our spectroscopic targets were selected using the Bayesian inference model of \citetalias{2021MNRAS.507.1127K}, which provided membership probabilities based on $\it Gaia$ photometry and astrometry. The measurement of line-of-sight velocities and [Fe/H] allowed us to assess which of these candidates were fully consistent with membership of $\omega$ Cen.  We find that the majority of our nominated members star  ($P_\text{mem}>0.3$) are indeed consistent with belonging to $\omega$ Cen, nominating 157 {\it bona fide} stars across the six degree of extent of our study. Five of these stars lie in the tidal extensions, while several more in the tails are nominally low-probability members but with line-of-sight velocities and [Fe/H] measurements strongly suggestive of $\omega$ Cen membership.

We explore the properties of the peripheral and tidal tail stars as a function of position along the direction of the stream. Due to the paucity of measurements at large distances along the tails, it is not possible to draw firm conclusions about the radial behaviour but we see no evidence of strong gradients in either $\textrm{V}_\textrm{h}$ or [Fe/H].  We compare our measurements to those in the literature as well as to the predictions of an N-body simulation of the cluster's disruption from \citetalias{2019NatAs.tmp..258I}. Our $\textrm{V}_\textrm{h}$ measurements show good consistency with those from \citetalias{2019NatAs.tmp..258I}, which was specifically tailored to match the properties of the \textit{Fimbulthul} stream. We also find good agreement with the measurements presented in  \citetalias{2012ApJ...751....6D} which cover the radial range of 10 - 40 arcmin, lying inward of and slightly overlapping with our data. Furthermore, our [Fe/H] abundances compare well to those of the $\omega$ Cen extra-tidal population presented in \citet[][]{2025MNRAS.537.2752K}, based on the Pristine photometric survey. We find a difference in metallicity dispersion compared to that study, but believe this is likely due to a sampling issue related to the target brightness of the two surveys.

While we have presented the most extensive spectroscopic study yet of the outer regions of $\omega$ Cen, our VLT/FLAMES fields still cover only a small area compared to the large extent of the cluster periphery and its tails.  Even with our Bayesian inference technique to guide us on the selection of probable members, the number of stars detected in $\omega$ Cen tidal tails remains very small. Much larger samples of stars at large radius will be required to fully understand $\omega$ Cen's origin, disruption and link to other substructures in the MW halo. The vast spatial extent and sparse nature of the tidal extensions demands a specialised approach, combining modeling work with deep wide-field multi-object spectroscopy. New and forthcoming facilities, such as WEAVE on the WHT 4.2m \citep{2012SPIE.8446E..0PD}, 4MOST on the VISTA 4m telecope \citep{2019Msngr.175....3D,2023Msngr.190...13L}, MOONs \citep[][]{2020Msngr.180...10C} and the Prime Focus Spectrograph \citep[PFS;][] {2016SPIE.9908E..1MT} are ideally suited to the study of GC outskirts. In the near future, we will be able to dissect the faint structures around $\omega$ Cen fully, and uncover the secrets of the tidal tails belonging to the most enigmatic cluster in the Galaxy.

\section*{Acknowledgements}
We thank the anonymous referee for their prompt attention and accurate comments. We thank Rodrigo Ibata for graciously supplying the simulated GC system for comparison. We thank Sergey Koposov for guidance in running \textsc{rvspecfit}. ALV and PBK acknowledge support from a UK Research and Innovation (UKRI) Future Leaders Fellowship (MR/S018859/1; MR/X011097/1). AMNF is supported by UKRI under the UK government’s Horizon Europe funding guarantee [grant number EP/Z534353/1] and by the Science and Technology Facilities Council [grant number ST/Y001281/1].
This work has made use of data from the European Space Agency (ESA) mission Gaia (https://www.cosmos.esa.int/gaia), processed by the Gaia Data Processing and Analysis Consortium (DPAC, https://www.cosmos.esa.int/web/gaia/dpac/consortium). Funding for the DPAC has been provided by national institutions, in particular the institutions participating in the Gaia Multilateral Agreement.

This work makes use of the following software packages: \textsc{astropy} \citep{2013A&A...558A..33A,2018AJ....156..123A},  \textsc{matplotlib} \citep{Hunter:2007}, \textsc{numpy} \citep{2011CSE....13b..22V}, \textsc{PyCosmic} \citep[][]{2012A&A...545A.137H}, \textsc{rvspecfit} \citep[][]{2011ApJ...736..146K,2019ascl.soft07013K},  \textsc{scipy} \citep{2020SciPy-NMeth}. 

For the purpose of open access, the author has applied a Creative Commons Attribution (CC BY) licence to any Author Accepted Manuscript version arising from this submission.

%%%%%%%%%%%%%%%%%%%%%%%%%%%%%%%%%%%%%%%%%%%%%%%%%%
\section*{Data Availability}
We have provided the full list of stars and their corresponding measurements as part of the publication, which are available as online material. The observations that underlie this paper can be retrieved from the ESO archive with programme ID 108.22MM.001 and PI Kuzma.


\bibliographystyle{mnras}
\bibliography{Library_bibtex} 
%%%%%%%%%%%%%%%%%%%%%%%%%%%%%%%%%%%%%%%%%%%%%%%%%%

%%%%%%%%%%%%%%%%% APPENDICES %%%%%%%%%%%%%%%%%%%%%

%\appendix

%%%%%%%%%%%%%%%%%%%%%%%%%%%%%%%%%%%%%%%%%%%%%%%%%%

% Don't change these lines
\bsp	% typesetting comment
\label{lastpage}
\end{document}